\PassOptionsToPackage{table}{xcolor}

\documentclass[draft]{agujournal}

\journalname{Geophysical Research Letters}

\begin{document}

%
%


\title{Comparing Jupiter interior structure models to \textit{Juno}
gravity measurements and the role of a dilute core}

%
%




\authors{S. M. Wahl\affil{1}, W. B. Hubbard\affil{2}, B. Militzer\affil{1,3}, 
T. Guillot\affil{4}, Y. Miguel\affil{4}, N. Movshovitz\affil{5,6} Y. Kaspi\affil{7}, R. Helled\affil{6,8},
D. Reese\affil{9}, E. Galanti\affil{7}, S. Levin\affil{10}, J.E. Connerney\affil{11}, S.J. Bolton\affil{12}}

\affiliation{1}{Department of Earth and Planetary Science, University of
California, Berkeley, CA, 94720, USA}
\affiliation{2}{Lunar and Planetary Laboratory, The University of
Arizona, Tucson, AZ 85721, USA}
\affiliation{3}{Department of Astronomy, University of
California, Berkeley, CA, 94720, USA}
\affiliation{4}{ Laboratoire Lagrange, UMR 7293, Universit\'e de Nice-Sophia Antipolis,
CNRS, Observatoire de la C\^ote dAzur, CS 34229, 06304 Nice
Cedex 4, France }
\affiliation{5}{Department of Astronomy and Astrophysics, University of California,
Santa Cruz, CA 95064, USA}
\affiliation{6}{Department of Geophysics, Atmospheric, and Planetary Sciences
Tel-Aviv University, Israel}
\affiliation{7}{Department of Earth and Planetary Sciences, Weizmann
Institute of Science, Rehovot, Israel.}
\affiliation{8}{Institute for Computational Sciences,
University of Zurich, Zurich, Switzerland}
\affiliation{9}{LESIA, Observatoire de Paris, PSL Research University, CNRS, Sorbonne Universits, UPMC Univ. Paris 06, Univ. Paris Diderot, Sorbonne Paris Cit, 5 place Jules Janssen, 92195 Meudon, France}
\affiliation{10}{JPL, Pasadena, CA, 91109, USA}
\affiliation{11}{NASA/GSFC, Greenbelt, MD, 20771, USA}
\affiliation{12}{SwRI, San Antonio, TX, 78238, USA}





\correspondingauthor{Sean M. Wahl}{swahl@berkeley.edu}




\begin{keypoints}
\item Precise gravity measurements allow better predictions of interior
    structure and core mass.
\item Juno's gravity measurements imply an increase in the abundance of heavy elements deep in the planet, at or inside its metallic region. 
\item The inferred structure includes a dilute core, expanded to a significant fraction of Jupiter's radius.
\end{keypoints}

%
%


\begin{abstract}

The \textit{Juno} spacecraft  has measured
Jupiter's low-order, even gravitational moments, $J_2$--$J_8$, to an
unprecedented precision, providing important constraints on the density
profile and core mass of the planet. Here we report on a selection of
interior models based on \textit{ab initio} computer simulations of
hydrogen-helium mixtures. We demonstrate that a dilute core, expanded to a
significant fraction of the planet's radius, is helpful in reconciling the
calculated $J_n$ with \textit{Juno}'s observations. Although model
predictions are strongly affected by the chosen equation of state, the
prediction of an enrichment of $Z$ in the deep, metallic envelope over that
in the shallow, molecular envelope holds. We estimate Jupiter's core to
contain an 7--25 Earth mass of heavy elements. We discuss the current
difficulties in reconciling measured $J_n$ with the equations of state, and
with theory for formation and evolution of the planet.
\end{abstract}

%
%

%


%
%
%
%

\section{Introduction} \label{sec:intro}

The \textit{Juno} spacecraft entered an orbit around Jupiter in July of 2016, and
since then has measured Jupiter's gravitational field to high precision
\citep{bolton2017}.  Here we present a preliminary suite of interior structure models
for comparison with the low order gravitational moments ($J_2$, $J_4$, $J_6$ and
$J_8$) measured by \textit{Juno} during its first two perijoves \citep{Folkner2017}. 

A well constrained interior structure is  a primary means of testing models for
the formation of the giant planets. The abundance and distribution of elements
heavier than helium  (subsequently referred to as ``heavy elements'') in the
planet is key in relating gravity measurements to formation processes. In the
canonical model for the formation of Jupiter, a dense core composed
$\sim$10~$M_\oplus$ (Earth masses) of rocky and icy material forms first,
followed by a period of rapid runaway accretion of nebular gas
\citep{Mizuno1978,Bodenheimer1986,Pollack1996}. Recent formation models suggest
that even in the core accretion scenario, the core can be small ($\sim$ 2
$M_\oplus$) or be diffused with the envelope
\citep{venturini2016,lozovsky2017}. If Jupiter formed by gravitational
instability, i.e., the collapse of a region of the disk under self-gravity
\citep{Boss1997}, there is no requirement for a core, although a core could
still form at a later stage \citep{helled2014}.  Even if the planet initially
formed with a distinct rock-ice core, at high pressures and temperatures these
core materials become soluble in liquid metallic hydrogen
\citep{Stevenson1985,wilson2012a,Wilson2012b,Wahl2013,Gonzalez2013}. As a
result, the core will erode and the heavy material will be redistributed
outward to some extent. In this study we consider the effect of such a dilute
core, in which the heavy elements have expanded to a significant fraction of
Jupiter's radius.

Significant progress has been made in understanding hydrogen-helium mixtures at
planetary conditions
\citep{saumon1995,Saumon2004,Vorberger2007,Militzer2008,Fortney2010,Nettelmann2012,militzer2013a,becker2013,militzer2016},
but interior model predictions are still sensitive to the hydrogen-helium
equation of state used \citep{hubbard2016,miguel2016}. 
In Section \ref{sec:barotropes} we describe the derivation of barotropes from a
hydrogen-helium equation of state based on \textit{ab-initio} materials
simulations \citep{militzer2013a,hubbard2016}, make comparisons to other
equations of states, and consider simple perturbations to better understand
their effect on the models. In Section \ref{sec:model} we describe details of
these models including a predicted layer of ongoing helium rain-out
\citep{stevenson1977a,stevenson1977b,Morales2009,Lorenzen2009,Wilson2010,morales2013},
with consideration of a dilute core in Section \ref{sec:dilute}. We then
describe the results of these models in terms of their calculated $J_n$
(Section \ref{sec:trends}) and heavy element mass and distribution (Section
\ref{sec:core_mass}). Finally, in Section \ref{sec:conclusion} we discuss these
results in relation to the present state of  measurements of, as well as theory
for the formation and evolution of Jupiter.

\section{Materials and Methods} \label{sec:methods}

\subsection{Barotropes} \label{sec:barotropes}
In this paper we consider interior density profiles in hydrostatic equilibrium,
\begin{equation} \nabla P = \rho \nabla U,     
    \label{eq:hydrostatic} \end{equation}
where $P$ is the pressure and $\rho$ is the mass density. In order to find a consistent
density profile, we use a barotrope $P(\rho)$ corresponding to isentropic profiles
constructed from various equations of state. 

Most of the results presented are based on  density functional theory molecular
dynamics (DFT-MD) simulations of hydrogen-helium mixtures from
\citet{militzer2013a} and \citet{militzer2013b} (MH13).  For densities below
those determined by the \textit{ab initio} simulations ($P<5$~GPa), we use the
\citet{saumon1995} equation of state (SCvH), which has been used extensively in
giant planet modeling. The benefits of this simulation technique lie in its
ability to determine the behavior of mixture through the metallization
transition, and to directly calculate entropy for the estimation of adiabtic
profiles. The barotropes are parameterized in
terms of helium and heavy element mass fraction $Y$ and $Z$, and specific
entropy $S$ as a proxy for the adiabatic temperature profile; for additional
details see Supplementary Section S1.

For comparison, we consider  models using the \textit{ab initio} equations of
state of hydrogen and helium calculated by \citet{becker2013}(REOS3) with the
procedure for estimating the entropy described by \citet{miguel2016}. Finally,
we also consider models using the SCvH EOS through the entire pressure range of
the planet.  Although the SCvH EOS does not fit the most recent data from
high-pressure shockwave experiments \citep{hubbard2016,miguel2016}, it is 
useful for comparison since it has been used to constrain Jupiter models
in the past \citep[e.g.][]{Saumon2004}.  

Different equations of state affect model outcomes in part by placing
constraints on the allowable abundance and distribution of heavy elements. The
DFT-MD isentrope consistent with the \textit{Galileo} probe measurements has
higher densities, and a less steep isentropic temperature profile than SCvH in
the vicinity of the metallization transition
\citep{militzer2013a,militzer2016}. The H-Reos equation of state has a similar
shape to the $T(P)$ profile, but has an offset in temperature of several
hundred K through much of the molecular envelope
\citep{Nettelmann2012,hubbard2016,miguel2016}. 

DFT-MD simulation is the best technique at present for determining
densities of hydrogen-helium mixtures over most of conditions in a giant planet
($P>5$ GPa).  There is, however, a poorly characterized uncertainty in density
for DFT-MD calculations. Shock-wave experiments are consistent with DFT, but
can only test their accuracy to, at best $\sim$6 \%
\citep{Knudson2004,Brygoo2015}.  Moreover, there is a necessary extrapolation
between $\sim$5 GPa, where the simulations become too computationally expensive
\citep{militzer2013a,militzer2013b}, and $\sim$10 bar where the deepest
temperature measurements from the \textit{Galileo} probe were obtained
\citep{Seiff1997}.  We consider perturbations to the MH13 equation of state in
the form of an entropy jump, $\Delta S$, at a prescribed pressure in the outer,
molecular envelope; increases of $S$ from 7.07 up to 7.30 (with S in units of
Boltzmann constant per electron) are considered. These perturbations test the
effect of a density decrease through the entire envelope ($P=$0.01 GPa), at the
switch from SCvH to DFT (5.0 GPa), and near the onset of the metalization
transition (50.0 GPa). 

Gravitational moments for the models are calculated using the non-perturbative
concentric Maclaurin spheroid (CMS) method
\citep{hubbard2012,hubbard2013,hubbard2016,wahl2016}; see Supplementary Section
S2 for additional details. 

\subsection{Model assumptions}\label{sec:model}

One of the most significant structural features of Jupiter's interior
arises from a pressure-induced immiscibility of hydrogen and helium, which allows for
rain-out of helium from the planet's exterior to interior
\citep{stevenson1977a,stevenson1977b}. \textit{Ab initio} simulations
\citep{Morales2009,Lorenzen2009,Wilson2010,morales2013} predict that the onset of this
immiscibility occurs around $\sim$100 GPa, over a similar pressure range as the
molecular to metallic transition in hydrogen. At higher pressures, the miscibility
gap closure temperature remains nearly constant with pressure, such that in the deep
interior temperatures are sufficient for helium to become miscible again.

The MH13 adiabats cross the \citet{morales2013} phase diagram such 
that helium rain-out occurs between $\sim$100-300 GPa \citep{militzer2016}.
This is consistent with the sub-solar $Y$ measurement made by the
\textit{Galileo} entry probe \citep{Zahn1998}. The REOS3 adiabats are
significantly warmer and require adjustments to the phase diagram in order to
explain the observations \citep{nettelmann2015}.  Although the detailed physics
involved with the formation and growth of a helium rain layer is poorly
understood \citep{Fortney2010}, the existence of a helium rain layer has a
number of important consequences for the thermal and compositional structure of
the planet.

We calculate the abundance of helium in both the upper helium-poor
(molecular hydrogen) region and lower helium-rich (metallic hydrogen) region by
enforcing a helium to hydrogen ratio that is globally protosolar. We also allow
for a compositional gradient of heavy elements across the layer with a mass
mixing ratio that changes from $Z_1$ in the lower layer to $Z_2$ in the upper
layer.

\subsection{Dilute Core} \label{sec:dilute}

The thermodynamic stability of various material phases in giant planet
interiors has been assessed using DFT-MD calculations
\citep{wilson2012a,Wilson2012b,Wahl2013,Gonzalez2013}.  These calculations
suggest that at the conditions at the center of Jupiter, all likely abundant
dense materials will dissolve into the metallic hydrogen-helium envelope. Thus,
a dense central core of Jupiter is expected to be presently eroded or eroding.
However, the redistribution of heavy elements amounts to a large gravitational
energy cost and the efficiency of that erosion is difficult to assess
\citep[see][]{Guillot2004}.  It was recently shown by \citet{vazan2016},  that
redistribution of heavy elements by convection is possible, unless the initial
composition gradient is very steep.  Some formation models suggest that a
gradual distribution of heavy elements is an expected outcome, following the
deposition of planetesimals in the gaseous envelope \citep{lozovsky2017}. The
formation of a compositional gradient could lead to double-diffusive convection
\citep{Chabrier2007,Leconte2013} in Jupiter's deep interior, which could lead
to a slow redistribution of heavy elements, even on planetary evolution
timescales.

\begin{figure}[h]
\centering

\includegraphics[width=20pc]{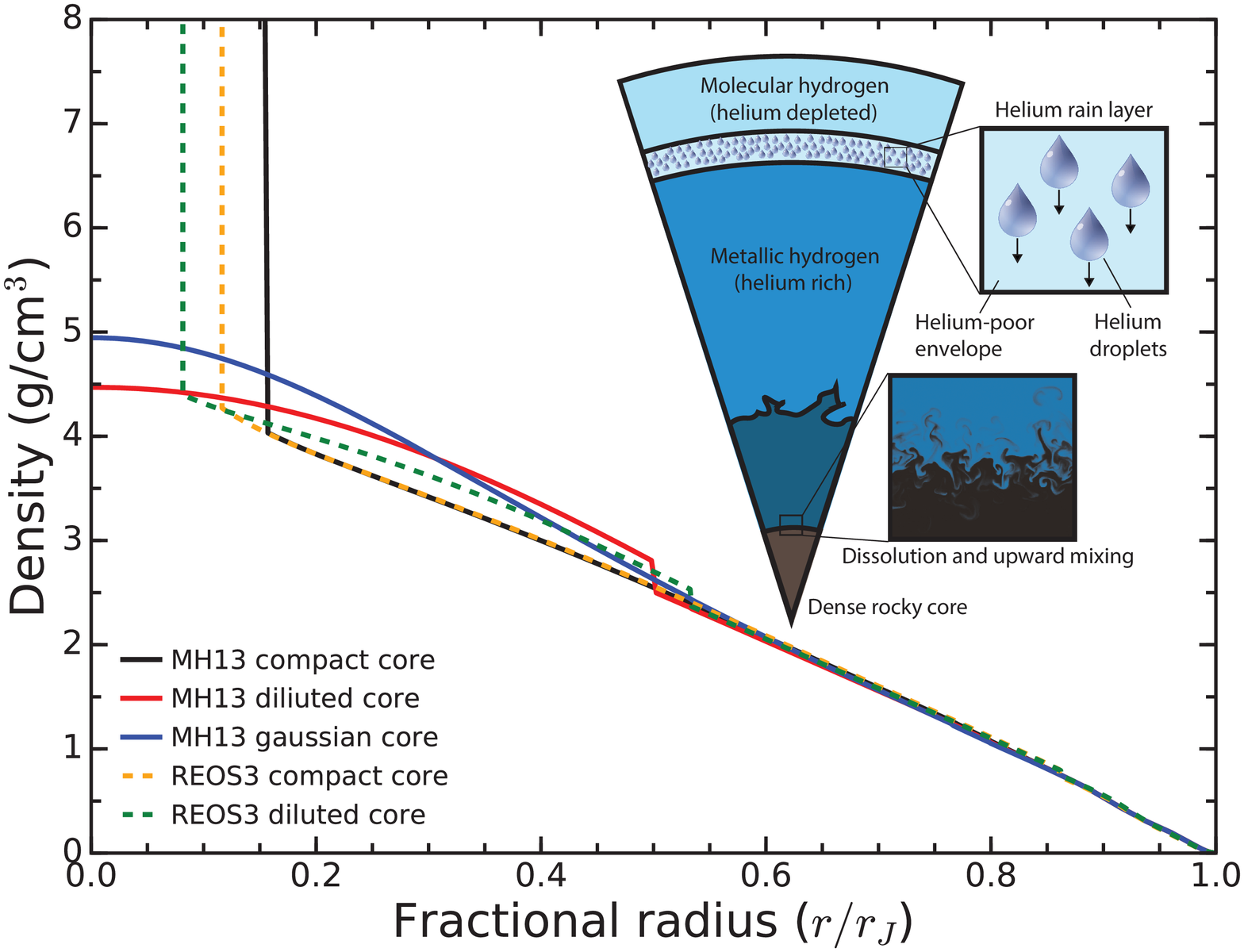}

\caption{Density profiles of representative models. Solid lines denote models
    using MH13, while dashed use REOS3. In black is a model with $S$, $Y$ and $Z$
    matching that measured by the \textit{Galileo} entry probe, and a core with constant
    enrichment of heavy elements inside $r/r_J$=$0.15$.  In red (Model
    D) $Z$=$0.007$ in the molecular envelope and constant $Z$-enriched, dilute
    core expanded to $r/r_J\sim0.50$ to fit the $J_4$ observed by
    \textit{Juno}. In blue (Model E) with $Z$=$0.007$ also fitting $J_4$ with
    Gaussian $Z$ profile. In orange (Model R) and green (Model S) are profiles
    for the REOS3 models fitting $J_4$ with a compact and dilute core,
    respectively. (Inset) Schematic diagram showing the approximate location of
the helium rain layer, and dilute core.}
\label{fig:density}
\end{figure}

In a selection of the models presented here, we consider Jupiter's ``core'' to
be a region of the planet in which $Z$ is enriched by a constant factor
compared to the envelope region exterior to it. This means that the model core
is a diffuse region composed largely of the hydrogen-helium mixture. In fact,
this configuration is not very different from the internal structure derived by
\citet{lozovsky2017} for proto-Jupiter.  Given the current uncertainty in the
evolution of a dilute core, we consider models with core in various degrees of
expansion, $0.15<r/r_J<0.6$.  In a few models, we also test the importance of
the particular shape of the dilute core profile by considering a core with a
Gaussian $Z$ profile instead.  Fig.~\ref{fig:density} demonstrates the density
profiles resulting from these different assumptions about the distribution of
core heavy elements. 

\section{Results} \label{sec:results}

\subsection{Comparison to \textit{Juno}} \label{sec:comparison}

The even zonal moments observed by \textit{Juno} after the first two perijoves
\citep{Folkner2017} are broadly consistent with the less precise predictions of
\citet{Campbell1985} and \citet{Jacobson2003}, but inconsistent with the more
recent JUP310 solution \citep{Jacobson2013}.  Table~\ref{tab:models} compares
these observations with a few representative models.  Although the solid-body
(static) contribution dominates this low-order, even part of the gravity
spectrum \citep{Hubbard1999}, a small dynamical contribution above
\textit{Juno}'s expected sensitivity must be considered \citep{Kaspi2010}.  For
sufficiently deep flows, these contributions could be many times larger than
\textit{Juno}'s formal uncertainties for $J_n$ \citep{Kaspi2017}, and thus
represent the conservative estimate of uncertainty for the purpose of
constraining the interior structure. Thus, ongoing gravity measurements by
\textit{Juno}, particularly of odd and high order, even $J_n$, will continue to
improve our understanding of Jupiter's deep interior \citep{Kaspi2013}.  Marked
in yellow in Fig~\ref{fig:j4j6}, is the possible uncertainty considering a wide
range of possible flows, and finding a corresponding density distribution
assuming the large scale flows are to leading order geostrophic
\citep{Kaspi2009}. The progressively smaller ellipses show this how this
    uncertainty is reduced when the depth of the flow is restricted 10000, 3000
and 1000 km respectively.  The relatively small range in our model $J_6$ and
$J_8$ compared to these uncertainties suggests flow in Jupiter are shallower
than the most extreme cases considered by \citet{Kaspi2017}.

\subsection{Model Trends} \label{sec:trends}

It is evident that the $J_n$ observed by \textit{Juno} are not consistent with
the ``preferred'' model put forward by \citet{hubbard2016}, even considering
differential rotation. Nonetheless, we begin with a similar model (Model A in
Tab.  \ref{tab:models}) since it is illustrative of the features of the model
using the MH13 equation of state with reasonable pre-\textit{Juno} estimates
for model parameters. A detailed description of the reference model is included
Supplementary Section S3.

In order to increase $J_4$ for a given planetary radius and $J_2$,  
one must either increase the density below the 100 GPa pressure level or
conversely decrease the density above that level \citep[][their
Fig.~5]{Guillot1999}. We explore two possibilities: either we raise the density
in the metallic region by expanding the central core, or we consider the
possibility of an increased entropy in the molecular region.

\begin{figure}[h]
\centering

\includegraphics[width=18pc]{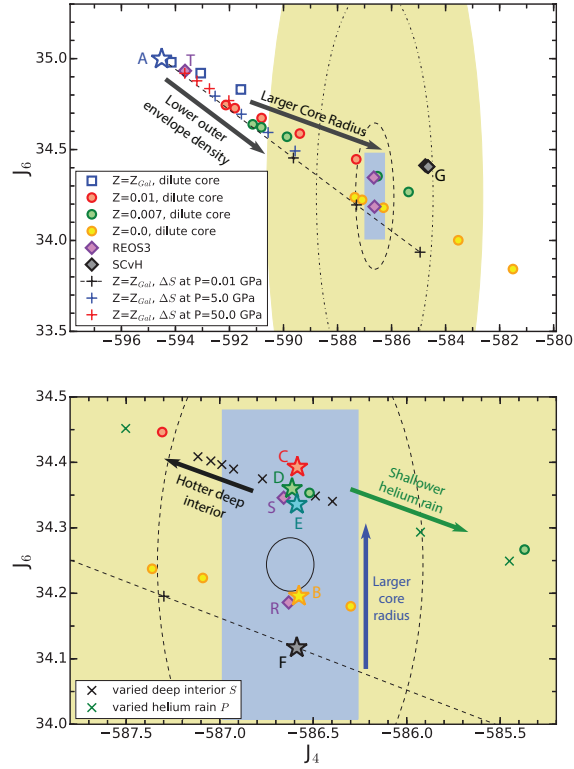}

\caption{Zonal gravitational moments $J_4$ and $J_6$ for interior models
    matching the measured $J_2$.
    (Upper) The blue rectangle shows the uncertainty of the \textit{Juno}
    measurements as of perijove 2 \citep{Folkner2017}. The yellow ellipse shows
    the effective uncertainty in the static contribution due possible deep
    differential rotation \citep{Kaspi2017}  and with flow restricted to 10000
    km (dash-dot), 3000 km (dashed), and 1000 km (solid). The blue star is the
    reference (Model A, Tab.~\ref{tab:models}) with $Z_1=Z_{\rm Gal}$ 
    matching that measured by the \textit{Galileo} entry probe, and an core of
    $r/r_J$=$0.15$. The blue squares show how these results change as a dilute
    core with a constant $Z_1$ enrichment and core radius $r$ increasing
    to the right. The green and red circles denote similar expanding core
    trends with lowered outer envelope heavy element fraction to 
    $Z_1$=$0.007$ and $Z_2$=$0.01$, respectively.  The `+'s denote models
    which take perturb the MH13 EOS by introducing a jump in $S$ at $P$=$0.01$
    (black), $P$=$5.0$ (blue) and $P$=$50.0$ GPa (red), with $Z_1$
    decreasing to the right. Black diamonds show models using the SCvH EOS.     
    (Lower) Models fitting the observed $J_4$ yield larger $J_6$ with increasing core
    radii. The stars denote models B, C, D, E, \& F in Table~\ref{tab:models}.
    Violet diamonds show models using the REOS3 EOS (Models R, S \& T).  Black
    and green `x's show models starting with the green star (dilute core,
    $Z_1$=$0.007$) and changing the $S$ of the deep interior or the
    pressure of the onset of helium rain. Red, green and cyan stars show
    models fitting the measured $J_4$ with the radius of the dilute core.
    Black Star shows model fitting $J_4$ with with the entropy jump magnitude
    $\Delta S$. 
}
\label{fig:j4j6}
\end{figure}

Fig.~\ref{fig:j4j6} shows the effect of increasing the radius of the dilute
core on $J_4$ and $J_6$. Starting with the MH13 reference model with
$r/r_J=0.15$ (Model A), the core radius is increased incrementally to
$r/r_J\sim0.4$, above which the model becomes unable to fit $J_2$.  Therefore,
considering an extended core shifts the higher order moments towards the
\textit{Juno} values, but is unable to reproduce $J_4$, even considering a
large dynamical contribution to $J_n$.  Supplementary Fig.~S1 shows a similar
trend for $J_8$, although the relative change in $J_8$ with model parameters
compared to the observed value is less significant than for $J_4$ and $J_6$. 

Precisely matching \textit{Juno}'s value for $J_4$ with the MH13 based models
presented here, requires lower densities than the reference model through at
least a portion of the outer, molecular envelope. In the absence of additional
constraints, this can be accomplished by lowering $Y$ or $Z$, or by increasing
$S$ (and consequently the temperature). In Fig.~\ref{fig:j4j6} this manifests
itself as a nearly linear trend in $J_4$ and $J_6$ (black `+' symbols), below
which there are no calculated points.  This trend also improves the agreement
of $J_4$ and $J_6$ with \textit{Juno} measurements, but with a steeper slope in
$J_6/J_4$ than that from the dilute core.  For $\Delta S\sim0.14$ applied at
$P=$0.01 GPa, a model with this perturbed equation of state can match the
observed $J_4$, with a mismatch in  $J_6$ of $\sim0.1\times10^{-6}$ below the
observed value (Model F).  When the $\Delta S$ perturbation is applied at
higher pressures ($P=5.0$ and $50.0$ GPa), a larger $\Delta S$ is needed to
produce the same change in $J_4$.

We also consider a number of models with both a decrease in the density of the outer,
molecular layer and a dilute core. Here we present MH13 models where the core
radius is increased for models with outer envelope $Z=0.010$, $0.007$ or $0.0$.
Above $Z\sim0.010$ the models are unable to simultaneously match $J_2$ and $J_4$. The
models with $Z=0.010$ and $Z=0.007$ can both fit $J_4$, but with a $J_6$ 
$\sim0.1\times 10^{-6}$ above the observed value (Models C \& D). These models also
require extremely dilute cores with $r/r_J\sim0.5$ in order to match $J_4$. A more
extreme model with no heavy elements ($Z=0$) included in the outer, molecular
envelope (Model B) can simultaneously match $J_4$ and $J_6$ within the current
uncertainty, with a less expansive core with $r/r_J\sim0.27$. The dilute
core using the Gaussian profile and an outer envelope $Z=0.007$ (Model E), has a very similar
trend in $J_4$--$J_6$, although it is shifted to slightly lower values of $J_6$.

There are a number of other model parameters which lead to similar, but less
pronounced, trends than the dilute core. Starting with Model C, we test
shifting the onset pressure for helium rain, between 50 to 200 GPa, and the
entropy in the deep interior, $S=7.07$ to $7.30$ (lower frame in Fig.
\ref{fig:j4j6}).  Both modifications exhibit a similar slope in $J_4$--$J_6$ to
the models with different core radii, but spanning a smaller range in $J_4$
than for the dilute core trend. 

The models using REOS3 have a significantly hotter adiabatic $T$ profile than
MH13.  Models R and S (\ref{tab:models}) are two example solutions obtained
with the REOS3 adiabat, for a 3-layer model with a compact core, and when
adding a dilute core, respectively.  Because of the flexibility due to the
larger $Z$ values that are required to fit Jupiter's mean density, there is a
wide range of solutions \citep{Nettelmann2012, miguel2016} with $J_4$ values
that can extend all the way from $-599\times 10^{-6}$ to $-586\times 10^{-6}$, spanning the
range of values of the MH13 solutions. Model T corresponds to a model calculated 
with the same $\Delta Z$ discontinuity at the molecular-metallic transition as Model S 
but with a compact instead of dilute core. This shows that, as in the case of the MH13 EOS, 
with all other parameters fixed, a dilute core yields larger $J_4$ values.  

For both DFT-based equations of state, we find that heavy element abundances
must increase in the planet's deep interior. The required $\Delta Z$ across the
helium rain layer is increased with the REOS3 equation of state, and
decreased by considering a dilute core.  Regardless of the EOS used, including
a diffuse core has a similar effect on $J_6$, increasing the value by a
similar amount for similar degree of expansion, when compared to an analogous
model with a compact core. Thus $J_6$ may prove to be a useful constraint in
assessing the degree of expansion of Jupiter's core. 

\subsection{Predicted Core Mass} \label{sec:core_mass}

\begin{figure}[h]
\centering

\includegraphics[width=20pc]{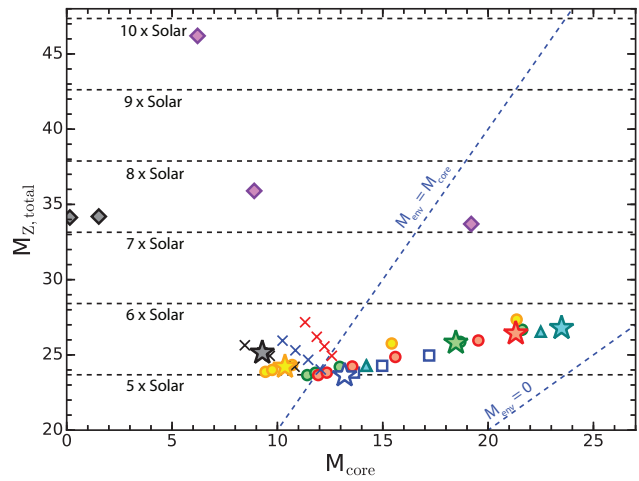}

\caption{Mass of heavy elements in the core of the model versus the total heavy
    element mass in Jupiter predicted by the model. Symbols refer to identical models
    as in Fig~\ref{fig:j4j6}. The stars denote models included in
    Table~\ref{tab:models}.  Horizontal lines display the values of $M_{\rm
    Z,total}$, corresponding to 5, 6, 7 and 8$\times$ solar abundance of heavy elements.
}
\label{fig:coremass}
\end{figure}

Fig. \ref{fig:coremass} displays the total mass of heavy elements, along with
the proportion of that mass in the dilute core. Models using MH13 with dilute
cores, have core masses between 10 and 24 ~$M_\oplus$ (Earth masses), with
gradual increase from 24 to 27~$M_\oplus$ for the total heavy elements in the
planet. Of the models able to fit the observed $J_4$, those with heavy element
contents closer to the \textit{Galileo} value have more extended cores
containing a greater mass of heavy elements. 

The perturbation of the equation of state with an entropy jump, has an opposite
effect on the predicted core mass with respect to the dilute core, despite the similar
effect on the calculated $J_n$. For increasingly large $\Delta S$ perturbations, core
mass decreases, to $\sim$8~$M_\oplus$, while total heavy element mass increases.  As
this perturbation is shifted to higher pressures the change in core mass becomes less
pronounced, for a given value of $\Delta Z$. In all the cases considered here, the
MH13 equation of state predicts significantly larger core masses and lower
total heavy element mass than the SCvH equation of state.

All of the models depicted in Fig. \ref{fig:coremass} represent fairly conservative
estimates of the heavy element mass. For any such model, there is a trade-off in
densities that can be introduced where the deep interior is considered to be hotter
(higher $S$), and that density deficit is balanced by a higher value of $Z$. 
It is also possible, that a dilute core would
introduce a superadiabatic temperature  profile, which would allow for a similar
trade-off in densities and additional mass in the dilute core.  Constraining this
requires an evolutionary model to constrain the density and temperature gradients
through the dilute core \citep{Leconte2012,Leconte2013}, and has not been
considered here.  Shifting the onset pressure of helium rain can shift the core mass
by $\sim$2~$M_\oplus$ in either direction. If the majority
of the heavy core material is denser rocky phase \citep{Soubiran2016}, the
corresponding smaller value of $\rho_0/\rho_Z$  results in a simultaneous decrease in
core mass and total $Z$ of $\sim$2--4~$M_\oplus$.

Using the REOS3, both models with a small, compact core of $\sim$6~$M_\oplus$ or a
diluted core of $\sim$19~$M_\oplus$ are possible, along with a continuum of
intermediate solutions.  These models have a much larger total mass of heavy
elements, $46$ and $34\,\rm M_\oplus$, a direct consequence of the higher
temperatures of that EOS \citep[see][]{miguel2016}.  The enrichment in heavy elements
over the solar value in the molecular envelope correspond to about $1$ for model R
and $1.4$ for model S, pointing to a water abundance close to the solar value in the
atmosphere of the planet.  In spite of the difference in total mass of heavy
elements, the relationship between core mass and radius is similar for MH13 and
REOS3.

In lieu of additional constraints we can likely bracket the core mass between
6--25~$M_\oplus$, with larger masses corresponding to a more dilute profile of
the core. These  masses for the dilute core are broadly consistent those
required by the core-collapse formation model \cite{Pollack1996}, as well as
models  that account for the dissolution of planetesimals \citep{lozovsky2017}.
The mass of heavy elements in the envelope, and thus the total heavy element
mass is strongly affected by the equation of state, with MH13 predicting
5--6$\times$ solar fraction of total heavy elements in Jupiter and REOS3 around 
$7-10$$\times$ solar fraction.

\section{Conclusion} \label{sec:conclusion}

After only two perijoves the \textit{Juno} gravity science experiment has
significantly improved the measurements of the low order, even gravitational
moments $J_2$--$J_8$ \citep{Folkner2017}. The formal uncertainty on these
measured $J_n$ is already sufficiently small that they would be able to
distinguish small differences between interior structure models, assuming that
the contribution to these low order moments arises primarily from the static
interior density profile. Considering a wide range of possible dynamical
contributions increases the effective uncertainty of the static $J_2$--$J_8$ by
orders of magnitude \citep{Kaspi2017}. It is expected that the dynamical
contribution to $J_n$ will be better constrained following future perijove
encounters by the \textit{Juno} spacecraft with measurements of odd and higher
order even $J_n$ \citep{Kaspi2013}. 

Even with this greater effective uncertainty, it is possible to rule out
a portion of the models presented in this study, primarily on the basis on the
observed $J_4$. The reference model, using a DFT-MD equation of state with
direct calculation of entropy in tandem with a consistent hydrogen-helium
phase-diagram is incompatible with a simple interior structure  model
constrained by composition and temperature from the \textit{Galileo} entry
probe. 

Our models suggest that a dilute core, expanded through a region 0.3--0.5
times the planet's radius is helpful for fitting the observed $J_n$.  Moreover,
for a given $J_4$ the degree to which the core is expanded affects $J_6$ and
$J_8$ in a predictable, model independent manner, such that further
constraining $J_6$ and $J_8$ may allow one to determine whether Jupiter's
gravity requires such a dilute core. Such a core might arise through erosion
of an initially compact rock-ice core, or through a differential rate of
planetesimal accretion during growth, although both present theoretical
challenges.

Using the REOS3 approach leads to a wider range of possibilities which include
solutions with the standard 3-layer model approach or assuming the presence of
a dilute core.  In any case, as for the MH13 solutions, the REOS3 solutions
require the abundance of heavy elements to increase in the deep envelope. This
indicates that Jupiter's envelope has not been completely mixed. 

The dilute core models presented here are preliminary with few key
    assumptions, which may be relaxed with future work. The first is the simple
    adiabatic temperature profile through the deep interior, in lieu of more
    consistent profiles in $T$ and $Z$.  Second, is the use of the ideal volume
    law, which does not necessarily remain a good assumption for the high $Z$
    in the planets core. Based on the range of models with different interior
    $S$ and $Z$ we expect more realistic treatments to have only a minor on
    calculated $J_n$, although changes in the predicted heavy element contents
on the order of a few $M_\oplus$ can be expected. In any case, these
assumptions have a smaller effect on model predictions than the differences in
EOS at present.

These results present a challenge for evolutionary modelling of Jupiter's deep
interior  \citep[e.g.][]{vazan2016,mankovich2016}.  The physical processes
involved with the formation and stability of a dilute core are not understood.
It strongly depends on the formation process of the planet and the mixing at
the early stages after formation, and also enters a hydrodynamical regime of
double diffusive convection where competing thermal and compositional gradients
can result in inefficient mixing of material \citep{Leconte2012,Mirouh2012}.
The timescale for the formation and evolution of such features, especially on
planetary length scales is still poorly understood.  In particular, it is not
known whether there would be enough convective energy to expand 10~$M_\oplus$
or more of material to 0.3 to 0.5$\times$ Jupiter's radii. It is also presently
unknown whether it is plausible to expand the core to this degree without fully
mixing the entire planet, and without resorting to extremely fortuitous choices
in parameters.  Since Jovian planets are expected to go through periods of
rapid cooling shortly after accretion \citep{Fortney2010}, if they are mostly
convective, it is likely that much of the evolution of a dilute core would
have to occur early on in the planet's history when the convective energy is
greatest. This presents a challenge for explaining interior models requiring a
large $\Delta Z$ across the helium rain layer, as such a layer would form after the
period of most intense mixing.

In our preliminary models, those able to fit $J_4$ have lower densities in
portions of the outer molecular envelope than MH13. This is achieved though
modifying abundances of helium and heavy elements to be lower than those measured by
the \textit{Galileo} entry probe, or invoking a hotter non-adiabatic
temperature profile. Some formation scenarios \citep[e.g.][]{Mousis2012} can
account for relatively low envelope ${\rm H}_2{\rm O}$ content ($\sim2\times$
solar), but our models would require even more extreme depletions for this
to be explained by composition alone.  Alternatively there might be an
overestimate of the density inherent to the DFT simulations of MH13 of the
order of $\sim$3\% for $P<100$

Interior models could, therefore, be improved through further theoretical and
experimental studies of hydrogen-helium mixtures, particularly in constraining
density in the pressure range below $\sim$100 GPa, where the models are most
sensitive to changes in the equation of state. More complicated equation of state
perturbations, including the onset and width of the metallization transition
\citep{Knudson2017} may be worth considering in future modelling efforts.
Similarly, the interior modeling effort will be aided by an independent
measurement of atmospheric ${\rm H}_2{\rm O}$ from \textit{Juno}'s microwave
radiometer (MWR) instrument \citep{helled2014b}.




%
%
%
%
%
%
%

%

\begin{sidewaystable}
\caption{Comparison of selected models to observed gravitational moments}
\label{tab:models}
\begin{tabular}{llrrrrrrrrrrrrr}
    \hline
    
    & Model~Description$^{a}$ &    $Z_1$$^{c}$ &        $Z_2$ &            $J_2$ &          $J_4$ &         $J_6$        &        $J_8$ &        $J_{10}$ &     $C/Ma^2$ &  $r_{\rm core}/r_J$ &     $M_{\rm core}$ &        $M_{Z,{\rm env}}$ &  $M_{Z,{\rm total}}$ &  $Z_{\rm global}$ \\
\hline
& \it{Juno}~observed$^{b}$                        & & & $14696.514$   & $-586.623$   &  $34.244$   & $-2.502$ &   &  &         &  &         &  &       \\
 &                                         & & & $\pm 0.272$   & $\pm 0.363$   &  $\pm 0.236$  & $\pm 0.311$ &  &  &         &  &         &  &       \\
\hline
A & MH13,~$Z_{\rm~Gal}$,~compact~core                &  0.0169  &  0.0298  &  14696.641  &  -594.511  &  34.998  &  -2.533  &  0.209  &  0.26391  &  0.150  &  13.2  &  10.5  &  23.6  &  0.0744  \\
\rowcolor{blue!15}
B & MH13,~dilute~core                   &  0.0000  &  0.0451  &  14696.641  &  -586.577  &  34.196  &  -2.457  &  0.202  &  0.26400  &  0.270  &  10.4  &  13.9  &  24.2  &  0.0762  \\
\rowcolor{blue!15}
C & MH13,~dilute~core                  &  0.0100  &  0.0114  &  14696.467  &  -586.613  &  34.360  &  -2.481  &  0.205  &  0.26396  &  0.498  &  18.5  &  7.3   &  25.8  &  0.0812  \\
\rowcolor{blue!15}
D & MH13,~dilute~core                 &  0.0071  &  0.0199  &  14696.641  &  -586.585  &  34.392  &  -2.486  &  0.205  &  0.26396  &  0.530  &  21.3  &  5.1   &  26.4  &  0.0831  \\
\rowcolor{blue!15}
E & MH13,~Gaussian~core                &  0.0071  &  0.0087  &  14696.467  &  -586.588  &  34.336  &  -2.479  &  0.204  &  0.26397  &  --     &  23.5  &  3.3   &  26.8  &  0.0843  \\
\rowcolor{blue!15}
F & Perturbed~MH13,~compact~core  &  0.0169  &  0.0526  &  14696.466  &  -586.588  &  34.117  &  -2.444  &  0.200  &  0.26400  &  0.150  &  9.3   &  15.9  &  25.1  &  0.0791  \\
G & SCvH,~compact~core                 &  0.0820  &  0.0916  & 14696.641 & -587.437 &  34.699  &  -2.541  &  0.212  &  0.26393  &  0.150  &  1.5   &  32.7 &  34.2  &  0.1076  \\
\rowcolor{blue!15}
R & REOS3,~compact~core & 0.0131 & 0.1516 &  14696.594 & -586.631 & 34.186 &  -2.457 & 0.202 & 0.26443 & 0.110 & 6.21 & 40.0 & 46.2 & 0.1454 \\
\rowcolor{blue!15}
S & REOS3,~dilute~core & 0.0209 & 0.0909 & 14696.755 & -586.658 & 34.346 & -2.480 & 0.204 & 0.26442 & 0.533 & 19.2 & 14.5 & 33.7 & 0.1061 \\
T & REOS3,~compact~core, low $J_4$ & 0.0293 & 0.0993 & 14696.381 & -593.646 & 34.933 & -2.529 & 0.209 & 0.26432 & 0.122 & 8.9 & 27.0 & 35.9 & 0.1129 \\
\hline
\multicolumn{15}{l}{$J_n$ in parts per million. Shaded rows are models match the \textit{Juno} observed $J_2$--$J_8$ within the current uncertainty.} \\
\multicolumn{15}{l}{$^{a}$Equation of state used, dilute or compact core, $Z_{\rm~Gal}$ denotes model with $Z_1$ matching \textit{Galileo} probe measurement. $^{b}$\citet{Folkner2017}.} \\
\multicolumn{15}{l}{$^{b}$\citet{Folkner2017}.} \\
\multicolumn{15}{l}{$^{c}$$Z_1$ denotes the heavy element fraction in molecular envelope, $Z_2$ denotes heavy element fraction in the metallic envelope, but exterior to the core. } 
\end{tabular}
\end{sidewaystable}

\acknowledgments
This work was supported by NASA's Juno project. SW and BM acknowledge the
support the National Science Foundation (Astronomy and Astrophysics Research
Grant 1412646). TG and YM acknowledge support from CNES. We acknowledge the
helpful input and discussion from Johnathan Lunine, David Stevenson, William
Folkner and the \textit{Juno} Interior Working Group.

\listofchanges

\end{document}


\supportinginfo{Comparing Jupiter interior structure models to \textit{Juno}
gravity measurements and the role of an expanded core}

\authors{S. M. Wahl\affil{1}, W. B. Hubbard\affil{2}, B. Militzer\affil{1,3}, 
T. Guillot\affil{4}, Y. Miguel\affil{4}, N. Movshovitz\affil{5,6} Y. Kaspi\affil{7}, R. Helled\affil{6,8},
D. Reese\affil{9}, E. Galanti\affil{7}, S. Levin\affil{10}, J.E. Connerney\affil{11}, S.J. Bolton\affil{12}}

\affiliation{1}{Department of Earth and Planetary Science, University of
California, Berkeley, CA, 94720, USA}
\affiliation{2}{Lunar and Planetary Laboratory, The University of
Arizona, Tucson, AZ 85721, USA}
\affiliation{3}{Department of Astronomy, University of
California, Berkeley, CA, 94720, USA}
\affiliation{4}{ Laboratoire Lagrange, UMR 7293, Universit\'e de Nice-Sophia Antipolis,
CNRS, Observatoire de la C\^ote dAzur, CS 34229, 06304 Nice
Cedex 4, France }
\affiliation{5}{Department of Astronomy and Astrophysics, University of California,
Santa Cruz, CA 95064, USA}
\affiliation{6}{Department of Geophysics, Atmospheric, and Planetary Sciences
Tel-Aviv University, Israel}
\affiliation{7}{Department of Earth and Planetary Sciences, Weizmann
Institute of Science, Rehovot, Israel.}
\affiliation{8}{Institute for Computational Sciences,
University of Zurich, Zurich, Switzerland}
\affiliation{9}{LESIA, Observatoire de Paris, PSL Research University, CNRS, Sorbonne Universits, UPMC Univ. Paris 06, Univ. Paris Diderot, Sorbonne Paris Cit, 5 place Jules Janssen, 92195 Meudon, France}
\affiliation{10}{JPL, Pasadena, CA, 91109, USA}
\affiliation{11}{NASA/GSFC, Greenbelt, MD, 20771, USA}
\affiliation{12}{SwRI, San Antonio, TX, 78238, USA}


\correspondingauthor{Sean M. Wahl}{swahl@berkeley.edu}

\newcommand{\beginsupplement}{%
        \setcounter{table}{0}
        \renewcommand{\thetable}{S\arabic{table}}%
        \setcounter{figure}{0}
        \renewcommand{\thefigure}{S\arabic{figure}}%
     }
\beginsupplement

%
%

\section*{Contents}
\begin{enumerate}
\item Text S1 to S3
\item Figure S1 
\end{enumerate}





\section*{S1. Equations of State}

The \textit{ab initio} simulations for MH13 were performed at a single, solar-like
helium mass fraction, $Y_0=0.245$. The precise abundance and distribution for both
helium and heavy element fractions are, \textit{a priori} unknown.  These are
quantified in terms of their local mass fractions, $Y$ and  $Z$. Our models consider
different proportions of both components by perturbing the densities using a relation
derived from the additive value law \citep{hubbard2016}. For the helium density we
use the pure helium end-member of SCvH. We assume a density ratio of heavy element to
hydrogen helium mixture,  $\rho_0 / \rho_Z$, of  $0.38$ for pressures below $100$
GPa, corresponding to heavy element composition measured by the \textit{Galileo}
entry probe \citep{Wong2004}, and $0.42$ for a solar fraction at higher pressures;
see discussion in \citet{hubbard2016}.  
The MH13 equation of state  uses density functional theory molecular dynamics
(DFT-MD) simulations in combination with a thermodynamic integration to find the
entropy of the simulated material. This allows one to directly characterize an
adiabat for the \textit{ab initio} equation of state as the $T(P)$ path in which the
simulated entropy per electron $S/k_B/N_e$ remains constant.  Here $k_B$ is
Boltzmann's constant and $N_e$ is the number of electrons. In the following
discussion, the term ``entropy'' and the symbol $S$ are used interchangeably to refer
to the particular adiabatic temperature profile through regions of the planet
presumed to be undergoing efficient convection. In this work, we assume that the
compositional perturbations have a negligible effect on the isentropic temperature
profile \citep{Soubiran2016}.

Models calculated with REOS3 followed the approach described by
\citet{miguel2016}: We fitted separately the core mass and composition
in heavy elements. The helium content of the molecular region was
fixed to the Galileo value while the increase in helium abundance in
the metallic region was calculated to reproduce the protosolar
value. The abundance of heavy elements was allowed to be different in
the molecular and metallic regions.   

\section{S2. Calculation of Gravitational Moments} \label{sec:cms}

The unprecedented precision of \textit{Juno}'s gravity measurements presents a
challenge, as they are more precise than the perturbative methods historically used
to calculate $J_n$ from an interior structure model, \citep[e.g.][]{zharkov1978}. For
the results presented here, we instead use the non-perturbative, \textit{concentric
Maclaurin spheroid} (CMS) method
\citep{hubbard2012,hubbard2013,hubbard2016,wahl2016}. In this method, the density
structure is parameterized by $N$ nested, constant-density spheroids and  the
gravitational field is calculated as a volume-integrated function of all of the
spheroids. The method uses an iterative approach to find the shape of each spheroid,
such that the surface of each follows an equipotential surface of the total effective
potential, $U$, from the planet's self-gravity and the rotation.  The result is a model
with a self-consistent shape, internal density distribution and gravitational field.
The method has been shown to be precise and efficient, and has been benchmarked
against an independent, non-perturbative method \citep{wisdom2016}.

The CMS models presented here parameterize the spheroid radii using progressively
smaller $\Delta r$ from deep to shallow. The outermost layer has a $\Delta r$ of 1 km in
thickness, which allows the model to resolve the density structure consistent with
$P=1$~bar at the outer surface. We use an axisymmetric version of the CMS method with
510 spheroids, and a spherical harmonic expansion up to order $n=16$.

\section*{S3. Reference Interior Model}

The reference model (model A) fixes parameters in the outer (molecular) envelope to those
measured by the \textit{Galileo} entry probe: $S=7.074$, $Y=0.2333$ and $Z=0.0169$.
It should be noted that the $Z$ from \textit{Galileo} is based on a measurement
showing sub-solar ratio of ${\rm H}_2{\rm O}$ to other ices (i.e. ${\rm CH}_4$ and
${\rm NH}_3$) \citep{Wong2004}. It has been hypothesized that the entry probe may
have descended through an anomalously dry region of Jupiter's atmosphere, in which
case this value of $Z$ may be an underestimate. The helium ratio of the deep
(metallic) envelope is chosen assuming that the \textit{Galileo} $Y$ was depleted
from a solar composition by helium rain , and the deep entropy is chosen as a
moderate enhancement across the helium rain layer, $S=7.13$.  An upper and lower
pressure of the helium rain layer are determined by finding where the two adiabatic
profiles for the inner and outer envelope intersect the \citep{morales2013} phase
diagram. This step is done self-consistently for all values of $S$, except in a few
extreme cases where the corresponding adiabat does not intersect the phase diagram. 

The interior structures of the REOS3 models presented here differ in the treatment of
the helium rain, assuming a 3-layer boundary with a sharp transition between the
molecular and metallic envelopes. The difference $J_6$ between the REOS3 model with
the compact core (model X) and the perturbed EOS (model F) can be attributed to this
structural difference. 

The MH13 models assume that the helium-rain layer is
superadiabatic, a natural consequence of inefficient convection
\citep{militzer2016}. In the case of the REOS3 models, because the adiabat is
significantly warmer, the presence of such a superadiabatic region has minor
quantitative consequences on the solutions and was not considered. In that
case, we used the approach described in \citet{miguel2016}.  













\begin{figure}[h]
\centering

\includegraphics[width=20pc]{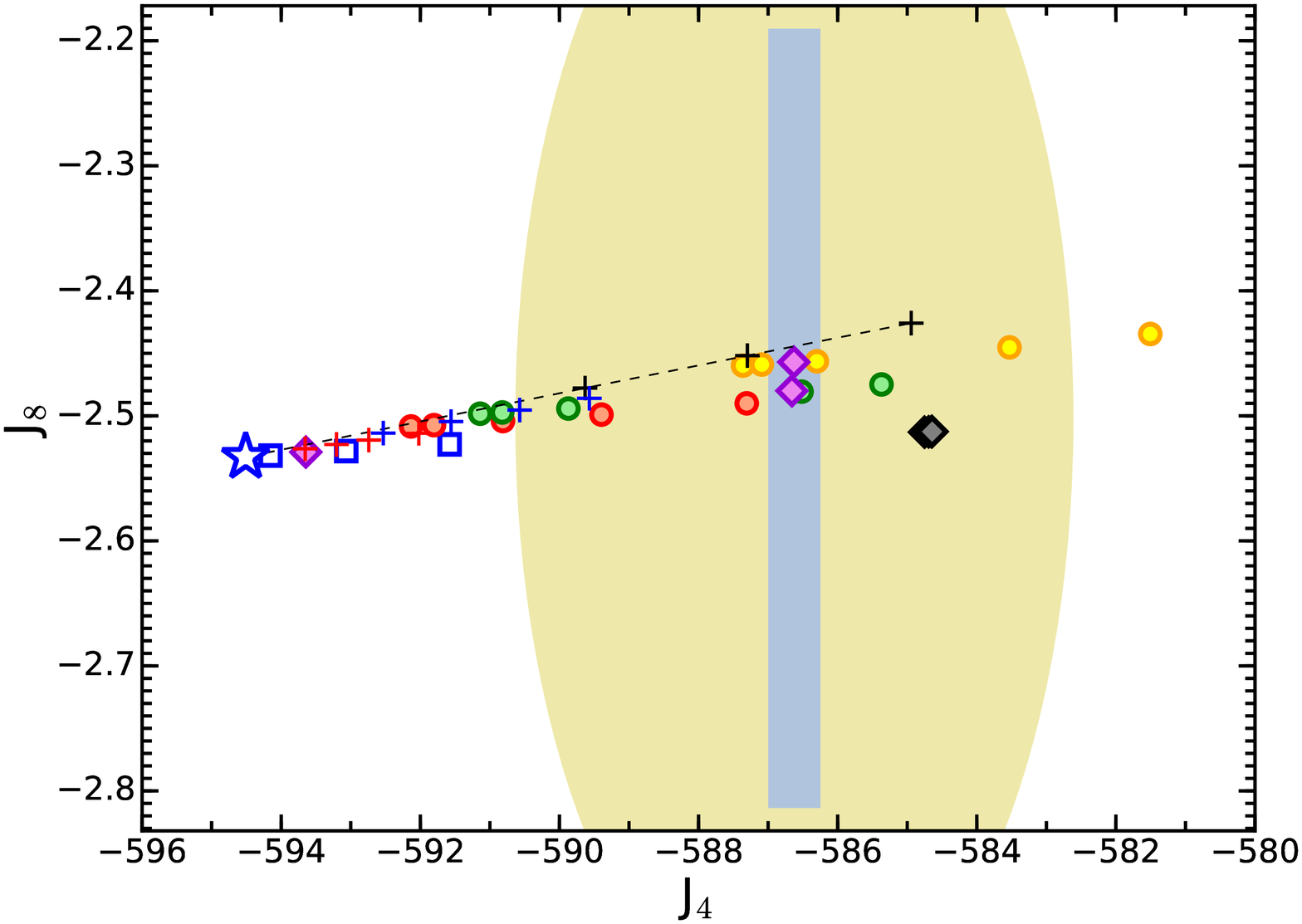}

\caption{ Zonal gravitational moments $J_4$ and $J_8$ for interior models
    matching the measured $J_2$. The rectangles show the uncertainty of the
    \textit{Juno} measurements as of perijove 2 \citep{Folkner2017}.The yellow region
    shows the effective uncertainty in the static contribution due possible deep
    differential rotation \citep{Kaspi2017}. Symbols refer to identical models as in
Fig. 2~in the main text. }
\label{fig:j4j8}
\end{figure}
